# Nonreciprocal Infrared Absorption via Resonant Magneto-optical Coupling to InAs

Komron J. Shayegan[1], Bo Zhao[2], Yonghwi Kim[1], Shanhui Fan[2], Harry A. Atwater[1*]

[1] *Thomas J. Watson Laboratory of Applied Physics, California Institute of Technology, Pasadena, California 91125*

[2] *Department of Electrical Engineering, Ginzton Laboratory, Stanford University, California 94305*

*Email*: shayegan@caltech.edu; bzhao8@Central.UH.edu; kim@ntt-research.com; shanhui@stanford.edu; haa@caltech.edu

## Abstract

Nonreciprocal elements are a vital building block of electrical and optical systems. In the infrared regime, there is a particular interest in structures that break reciprocity because their thermal absorptive (and emissive) properties should not obey the Kirchhoff thermal radiation law. In this work, we break time-reversal symmetry and reciprocity in *n*-type doped magneto-optic InAs with a static magnetic field where light coupling is mediated by a guided-mode-resonator (GMR) structure whose resonant frequency coincides with the epsilon-near-zero (ENZ) resonance of the doped InAs. Using this structure, we observe the nonreciprocal absorptive behavior as a function of magnetic field and scattering angle in the infrared. Accounting for resonant and nonresonant optical scattering, we reliably model experimental results that break reciprocal absorption relations in the infrared. The ability to design such nonreciprocal absorbers opens an avenue to explore devices with unequal absorptivity and emissivity in specific channels.

## Introduction

There has been great interest and numerous theoretical proposals for nonreciprocal absorbers on the grounds that they do not obey the Kirchhoff thermal radiation law (*1, 2, 3, 4*). Interest in this phenomenon stems from fundamental considerations, as well as applications ranging from infrared spectroscopy, sensing, and thermal energy conversion efficiency (*5, 6*). For infrared sensing applications, the amount of light absorbed in a given channel is something that could be controlled in a nonreciprocal device, where the absorptivity can be tuned from strongly in one channel and weak in another and vice-versa (*7, 8*). As we review in the coming section, nonreciprocal absorption achieved through an external magnetic field necessitates that the emission is channeled through the opposite channel. Nonreciprocal absorption is a required functionality to enable thermal energy harvesting exceeding the conversion efficiency constrained by the Shockley-Queisser limit (*9*) and reaching higher theoretical efficiency limits, such as the Landsberg limit (*10, 11, 12*).

The Kirchhoff thermal radiation law can be expressed as an equality of the absorptivity and emissivity at a given wavelength ($\lambda$), polarization, and angle of incidence ($\theta_i$) (*13, 14, 15*):

$$\alpha(\theta_i, \lambda) = e(\theta_i, \lambda) \tag{E1}$$

This equality between the angular and spectral distributions of the emissivity and absorptivity is a direct consequence of reciprocity, which in a scattering context takes the form:

$$r(\theta_i, \lambda) = r(-\theta_i, \lambda) \tag{E2}$$

where $r(\theta_i, \lambda)$ is the incident radiation from $\theta_i$ that is not absorbed by the absorber/emitter and reflected through the $-\theta_i$ channel and vice-versa for $r(-\theta_i, \lambda)$. The above relationships, however, assume that the emitter/absorber obeys Lorentz reciprocity and does not transmit any of the incident radiation (*16*). In a nonreciprocal system, the equality in equation E2 is broken; and the Kirchhoff law is violated (*17*). The nonreciprocal behavior of the reflection directly relates to the nonreciprocal thermal radiation. This relation is visualized in Figures 1A, B.

Achievement of nonreciprocity for infrared radiation is a subject of widespread investigation. Reciprocity can be broken with linear time-invariant, nonlinear, and linear time-varying platforms (*18, 19, 20, 21*). The use of magneto-optical materials, e.g., ferrites, in basic optical elements such as isolators and circulators has been a fundamental building block in integrated photonics for many decades. The use of bare magneto-optical materials to observe free-space radiation from coupled plasmon-phonon modes has been demonstrated, however the wavelength regimes are limited to the far-IR and the dependence on magnetic field was not measured (*22*). Prior experimental work on free-space magneto-optical elements has investigated coupling to surface plasmon resonances at a fixed incidence angle in the Otto configuration in the far-infrared regime (*23*), as well as polarization rotation in the Faraday geometry (*24, 25*). Pairing magneto-optical semiconductors with photonic crystals and examining the magnetic field and angular dependence in the infrared has not yet been experimentally explored. Furthermore, to date there has not been a measurement of the interaction of thermal radiation with materials that break reciprocal relations in the far and mid-infrared, despite an abundance of theoretical designs reporting free-space propagation with magneto-optical elements in this wavelength regime (*26, 27, 28*). Here we experimentally demonstrate strong nonreciprocal behavior, breaking the reflectivity relationship in equation E2 over a wide range of incident angles in an infrared thermal photonic absorber that incorporates magneto-optic InAs with a static external magnetic field (*29*). To apply the external magnetic field, we use a Halbach array consisting of three permanent magnets with their fields oriented towards a pole-piece (Figure 1C) (*30*). The pole-piece is made of a soft ferromagnetic alloy which focuses the field so that it is uniform over the sample. We control the strength of the field through the sample by tuning the length of the gap, *l*, between the two pole-pieces (Figure 1C, D). The maximum field used in this paper is limited by the sample size, 5 mm.

## Results

### Experiment

Our device consists of a low-loss, amorphous silicon (α-Si) GMR waveguide structure fabricated on the surface of a 500-μm-thick, degenerately *n*-type doped InAs wafer (Figures 1D, E) serving as our magneto-optical material resonantly excited in the Voigt geometry. The patterned α-Si structure on the InAs gives a clear angle-dependent guided mode dispersion relation with distinct resonances for both the zero and non-zero magnetic field cases. While most of the modal field of a guided mode is confined within a low-loss α-Si resonator, some of the modal field penetrates into the magneto-optic InAs layer below, which produces the nonreciprocal effect in the presence of a magnetic field (Figure 1F).

### Optical characterization of InAs and resonant infrared device design

The doping concentration of the InAs is chosen such that the GMR wavelength coincides with the ENZ wavelength (17.3 μm) in the infrared. The degenerately-doped InAs exhibits a Drude-like optical response with non-zero off-diagonal permittivity values in an applied magnetic field:

$$\varepsilon(\omega) = \begin{bmatrix} \varepsilon_{xx} & \varepsilon_{xy} & 0 \\ \varepsilon_{yx} & \varepsilon_{yy} & 0 \\ 0 & 0 & \varepsilon_{zz} \end{bmatrix}$$

(E3)

$$\varepsilon_{xx} = \varepsilon_{yy} = \varepsilon_{\infty} - \frac{\omega_p^2(\omega + i\Gamma)}{\omega[(\omega + i\Gamma)^2 - \omega_c^2]}$$

(E4)

$$\varepsilon_{xy} = -\varepsilon_{yx} = i\frac{\omega_p^2\omega_c}{\omega[(\omega + i\Gamma)^2 - \omega_c^2]}$$

(E5)

$$\varepsilon_{zz} = \varepsilon_{\infty} - \frac{\omega_p^2}{\omega(\omega + i\Gamma)}$$

(E6)

In equations E4-E6, the plasma and cyclotron frequencies are $\omega_p = \sqrt{ne^2/(m_e\varepsilon_0)}$ and $\omega_c = eB/m_e$, respectively. Four-point probe measurements of the InAs wafer in conjunction with ellipsometry fits of the values $\Psi$ and $\Delta$, the complex ratio amplitude and phase responses, give a relaxation rate $\Gamma = 4.5$ THz, a high-frequency limit dielectric constant $\varepsilon_{\infty} = 12.3$, a carrier concentration $n = 1.5 \times 10^{18}$ cm$^{-3}$, and an effective electron mass $m = 0.033\ m_e$, where $m_e = 9.109 \times 10^{-31}$ kg (*31*). The Drude model fit to the experimental values is shown in Figure 2A. Furthermore, our designs targeted the ENZ wavelength regime (17.3 µm) since the off-diagonal terms of the dielectric function ($\varepsilon_{xy}, \varepsilon_{yx}$) are largest relative to the on-diagonal terms when a magnetic field is applied to the InAs at this wavelength (Figure 2B).

We focused on controlling the nonreciprocal response around 17.3 µm since it corresponds to the portion of the electromagnetic spectrum exhibiting thermal radiation at or below room temperature. We implemented this control with a GMR structure, which is commonly employed to enhance absorption/reflection by critical coupling in infrared applications (*32, 33, 34, 35*). We designed the GMR structure to critically couple to *p*-polarized free-space radiation, (i.e., electric field oscillating in the *x-y* plane), near the ENZ wavelength over a range of angles, with the guided resonance in the thermal emitter. Because the resonator has been designed around the ENZ wavelength, we see a strong detuning of a resonant peak with a magnetic field. If no patterned α-Si were used, the plasma edge of the InAs would have no distinct resonance that could be detuned with a magnetic field. This effect is discussed in more detail in the following section.

The optimal dimensions for the GMR structure are found based on an analysis of the guided mode in a *uniform* α-Si slab atop *n*-InAs. In this approximation of the periodic α-Si, the uniform α-Si has a thickness $t = d_1/2 + d_2$, where $d_1$ is the periodic element depth and $d_2$ is the supporting layer thickness of the α-Si. The dispersion relation, $\omega(k_x)$, of the fundamental guided mode is then found by solving the equation (*26, 36, 37*):

$$\tan(k_{Si}t) = \frac{k_{Si}\varepsilon_{Si}[k_{air} + k_{InAs}\xi_{xx}]}{k_{Si}^2 - \varepsilon_{Si}^2 k_{air}k_{InAs}\xi_{xx}}$$

(E7)

where $k_{Si} = \sqrt{Re(\varepsilon_{Si})(\omega/c)^2 - k_x^2}$, $k_{air} = \sqrt{k_x^2 - (\omega/c)^2}$, $k_{InAs} = \sqrt{k_x^2 - (\omega/c)^2 Re(\varepsilon_{InAs})}$, $\xi_{xx} = 1/Re(\varepsilon_{InAs})$, and $k_x$ is the wavevector along the direction of propagation. In this derivation, we focus on the reciprocal case to get a sense of matching the slab waveguide to the Drude-like resonance of the InAs, and neglect the off-diagonal terms of the InAs which are zero when there is no applied magnetic field. In reference 26, the derivation for the non-zero magnetic field case shows that $\omega(k_x) \neq \omega(-k_x)$ and the symmetry of the dispersion relation is broken.

A plot of the symmetric angular dispersion for varying slab thicknesses of the α-Si waveguide is provided in the Supplementary Materials (Figure S1). In the calculations, we used the fitted dielectric constant values of the *n*-InAs from ellipsometry.

We chose a periodicity ($\Lambda$) of 7 µm such that the guided mode can be folded in the light cone and couple with the free-space radiation. We fine-tuned the other dimensions and used a GMR structure depth $d_1 = 0.72$ µm, and supporting layer thickness $d_2 = 2.1$ µm. Since the periodicity of the GMR structure itself is much smaller than the wavelength of interest, we expect to see only specular reflection in the considered wavelength range (*38*). We note that the linewidth of the resonance created by the GMR structure is sensitive to both the patterned area widths ($\Lambda/2$) and the scattering rate of the *n*-InAs underneath (*39*).

**Magnetic-field-dependent absorption tuning in magneto-optic GMR structure**

The Voigt geometry used in these experiments means that the nonreciprocal behavior of electromagnetic radiation inside of the InAs will be confined to $p$-polarized scattering, where the electric field oscillates perpendicular to the applied magnetic field (Figure 3A). Conversely, we expect reciprocal behavior for $s$-polarized scattering, as the applied magnetic field is aligned with the oscillations in the electric field, and the cross product of the electric and magnetic fields is zero. The reciprocal behavior for $s$-polarized light is shown in Figure 3B, where the absorptivity maxima for all magnetic field values are marked with a dashed gray line at approximately 17 μm. In Figures 3A and B, we emphasize the spectral shift in the peak position of the absorptivity and offset the data for clarity. The overall absorptivity for $s$-polarized light does not change with varying magnetic field. Furthermore, we do not observe any detectable polarization conversion from $p$ to $s$ or $s$ to $p$-polarization, confirming the proper alignment of our GMR structure relative to the optical beam path of the light (Figures 3C, D). The InAs layer is optically thick and therefor no light is transmitted through the sample, allowing us to only measure the absorptivity and reflectivity. We observed a significant redshift of the absorptivity peak due to the GMR, especially in the positive field case (0 T to 1.2 T). The strong magnetic dependence of the absorptivity spectrum clearly demonstrates the reciprocity breaking effect. This nonreciprocal effect is observed in a wide angular and wavelength range.

To deconvolve the contribution coming from the $n$-type InAs and the entire GMR structure ($n$-type InAs with α-Si on top), we simulated angular absorption spectra as a function of magnetic field for both cases and compared the simulations to our measurements (Figure 4). The simulations were performed using a finite-difference, frequency-domain electromagnetic simulation tool, and incorporated the measured InAs optical properties and α-Si grating parameters. The simulated and measured plasma edge splitting for the unpatterned InAs wafer are in good qualitative agreement, with a broad resonance visible for the positive field surface magneto-plasmon and a shorter wavelength onset (16.5 μm) for the negative field surface magneto-plasmon (Figures 4A, C). The addition of the periodic α-Si does not increase the wavelength shift induced by the plasma-edge splitting into two separate surface magneto-plasmons observed in the unpatterned InAs case. However, clear maxima are seen from the coupled GMR structure that can be evaluated for multiple angles and compared to the data, which is further discussed in the following section.

Comparing the measured positive field for the GMR structure (Figure 4B, red curve) and the simulation (Figure 4D, red curve) we note that both maxima occur at a resonance centered at 17.5 μm. However, a second shoulder resonance is present in the simulation at 16.7 μm which is not easily visible in the experiment. The shoulder resonance is much weaker in the experiment likely because of the larger material loss in both the α-Si and InAs. Going from zero to negative applied magnetic field at a $\theta_i = 50°$ we did not see a large shift in the resonant peak position for either the simulation or the measurement, but do see the resonance narrowing for the negative field case. The simulated and experimental traces for more angles of incidence and an intermediary 0.8 T field are included in the Supplementary Material (Figures S2-4).

To further highlight the effect of the GMR structure on nonreciprocal absorption, we look at the intensity effect on the absorption resonance at more oblique angles. Taking $\theta_i = 70°$, we can subtract the absorption of the bare InAs around the plasma shoulder from the GMR structure absorption and fit a Lorentzian to the difference (Figure 5A). When the magnetic field is then turned on, the resonant intensity peak can be drastically tuned from absorption that greatly exceeds the plasma shoulder (positive field) to well below (negative field).

**Angle-dependence of nonreciprocal absorption behavior**

To more comprehensively understand and characterize the nonreciprocal response of the hybrid GMR structure, we compare the experimental and simulated absorptivity maxima for $p$-polarized light over a range of incidence angles for $B = 0$ T, 0.8 T, and 1.2 T. To map our dispersion across positive and negative angles, we confirm that Onsager-Casimir relations hold (Figure S2). When $B = 0$ T, reciprocity is preserved, and we do not expect to see any reciprocity-breaking behavior. Consequently, the angular dispersion relation of the guided mode resonance at the plasma edge of the InAs remains symmetric. Figure 6A shows the symmetric case measured in our experiment and the corresponding absorptivity maxima from simulations (Figure 6D). We also observe a narrowing of the resonant linewidth for more oblique angles of incidence in both the simulated and experimental data when no magnetic field is applied, as shown by the vertical bars marking the full-width-at-5% maximum of each absorptivity spectrum. When the in-plane magnetic field is applied, reciprocity is broken, and the measured dispersion of the absorptivity maxima becomes asymmetric (Figures 6B, C). The degree of reciprocity breaking grows at smaller angles (approaching $\theta_i = 45°$), which is expected from the absorptivity maxima found from the simulations (Figure 6E, F).

We also observe a clear detuning of the absorptivity maxima with an increase in the magnetic field. For both the experimental (Figure 6B) and simulated (Figure 6E) absorptivity maxima at the intermediate 0.8 T magnetic field, the lineshape width for positive angles of incidence grows considerably. This makes the detuning of the absorptivity maxima at narrow incidence angles difficult to resolve. Increasing the magnetic field to 1.2 T separates the maxima at narrow angles such that the 5% bars do not overlap.

Unfortunately, our measurement system only allows for incidence angles down to 35°, limiting the ability to compare experiments and simulations for near-normal incidence. However, simulations indicate that the difference for positive and negative incidence (positive and negative field) decrease and approach zero as the scattering geometry approaches normal incidence since the first-order magneto-optic effects go to zero (Figure 6G-I).

Interestingly, the measured maxima for negative incidence angles do not shift below the zero magnetic field maxima for oblique angles. This contrasts with the maxima extracted from simulations, which redshift for oblique (-65° and -70°) angles of incidence.

## Discussion

In this study, we use a degenerately-doped magneto-optical semiconductor paired with a GMR coupled to free-space incident radiation to demonstrate nonreciprocal absorption in the infrared regime when a moderate magnetic field ($B$ up to 1.2 T) bias is applied. The degree of reciprocity breaking is largest at narrower incidence angles $\theta_i$ (45° - 55°), making the design potentially useful for cascading multiple emitters/absorbers to achieve directional flow of energy (40). The methods used to model, design, and fabricate the structure presented in this manuscript can be utilized for future implementations of nonreciprocal absorbers. As a future direction for engineering higher efficiency thermal radiators, direct measurement of the emissivity could be employed to demonstrate applications of systems that violate the Kirchhoff thermal radiation law. Higher quality factors of the structure could be obtained using a thinner magneto-optic layer with a back reflector or a lower doping concentration of InAs; however, this would result in a longer working wavelength. A higher mobility magneto-optic material, such as InSb, could also be used to achieve a narrower linewidth (41). The easily identifiable absorptivity peaks at the plasma edge are a consequence of using a periodic α-Si structure supporting a GMR, while the nonreciprocal behavior comes from the plasma-edge splitting of the magneto-optic $n$-type InAs. The temperature dependence of the optical properties of the $n$-InAs are included in the Supplementary

Material as a reference for designing at elevated temperatures to measure emissivity (Figure S5). Beyond demonstrating nonreciprocal absorption in the infrared, hybrid magneto-optic and photonic crystal structures like those reported here are of potential interest for thermal radiation control and free-space information processing owing to the polarization-dependent response and low required magnetic fields for resonance tunability.

## Materials and Methods

### Device Fabrication

Our fabrication of the GMR structure started by deposition of α-Si on the InAs wafer using plasma-enhanced chemical vapor deposition (PECVD). The deposition was carried out at a temperature of 200 °C and pressure of 800 mTorr with a flow rate of 250 SCCM (5% SiH$_4$/Ar) for 90 seconds. We then spin-coated 500 nm of ZEP 520A resist (1 minute at 2,000 rpm) and baked the sample for 5 minutes at 180 °C on a hot plate.

Electron beam lithography was then used to write the desired pattern into the resist, using a beam current of 100 nA with a 300 μm aperture and a dose of 240 μC/cm$^2$. After writing the pattern into the resist, the sample was dipped in ZED N-50 for 2 minutes and 30 seconds for development. Following exposure, the sample was baked at 140 °C for 3 minutes.

The α-Si pattern was subsequently etched using inductively coupled plasma-enhanced reactive ion etching (ICP RIE). The etching recipe began with an O$_2$ followed by a SF$_6$ cleaning cycle for 10 minutes each followed by 2 minutes and 30 seconds of etching with SF$_6$ as the etchant gas. The same cleaning cycle was then repeated in reverse. The sample was left in PG remover overnight and then checked with a confocal microscope to ensure the resist had been removed.

SEM images were then taken at normal and 45° tilted incidence to confirm the GMR structure dimensions. For the SEM images shown in this paper, the microscope was operated at an accelerating voltage of 5 kV.

### Measurements

To probe the sample, we used a two-theta stage with a silicon carbide Globar source and focusing optics, exciting the sample at an incident angle $\theta_i$ relative to the $x$-$z$ plane and collect the specular reflected light at $-\theta_i$. For the zero and low (up to but not including 0.8 T) magnetic field measurements, the original sample holder of the system (J. A. Woollam I.R. VASE Mark II) was used. For the higher magnetic field measurements, we used a separate adapter to accommodate the Halbach array. The lower magnetic field measurements are obtained using a specially-designed aluminum holder to separate the two neodymium magnets while holding the sample in between the two magnets' poles. The field strength was tuned by inserting aluminum spacers to increase the gap between the magnets. We used a Halbach array magnet configuration with tunable pole pieces to apply $0.8 < B < 1.2$ T (*42*). An image of the magnet assembly can be found in the Supplementary Material, Figure S7. We measured the magnetic field strength and uniformity across the sample surface with a transverse Hall sensor made from InAs (Lakeshore HGT 1010).

The sample was in the Voigt geometry, with an in-plane magnetic field, $B$, breaking time-reversal symmetry. By changing the sign of the magnetic field, we produced the same effect as keeping the sign of magnetic field constant while switching the positions of the source and detector (*43, 44, 45*). The use of linear polarizers at both the source output and detector input allow for a deconvolution of the physical effects (e.g. ensure that there is no cross-polarization due to sample misalignment).

**Device simulations**

We used the COMSOL *electromagnetic waves, frequency domain* package to simulate the optical response of the GMR structure. For the material properties and geometries of the structure, we used the measured values of the α-Si periodicity and depth from SEM images and *n*-InAs optical properties from the model fits of ellipsometry data and four-point probe measurements. For each trace at a fixed angle of incidence, we swept the wavelength from 16 to 19 μm in 12-nm intervals. The absorptivity was found by subtracting the reflectivity from 1, as there was no transmission for the structure. This was verified by taking transmission measurements of the *n*-InAs wafer.

**Acknowledgments**
**General:**

YK used the Kavli Nanoscience Institute (KNI) at Caltech for fabrication facilities. KS would like to acknowledge Hamid Akbari, Dennis Howard Drew, and Arne Laucht for discussions around the design of the experiment and application of the magnetic field.

**Funding:**

This work was supported by DARPA NLM (KS, BZ, YK, SF, HAA). KS would like to thank the support by the NSF GRFP for a Graduate Research Fellowship.

**Author contributions:**

The manuscript was written through contributions of all authors. All authors have given approval to the final version of the manuscript. Measurements were done by KS. Fabrication was done by YK. The theoretical and modeling work was led by BZ and supported by KS.

**Competing interests:** The authors declare that they have no competing interests.

**Data and materials availability:** All data discussed in this manuscript is presented and the analysis described. Additional data is available upon reasonable request.

## Figures and Tables

Fig. 1: Overview of nonreciprocal absorber/emitter theory, design, and implementation.
Fig. 2: Drude reflectance and dielectric constant of degenerately doped $n$-InAs.
Fig. 3: Polarization dependent reflection measurements for varying applied magnetic field.
Fig. 4: Magnetic field effect on $p$-polarized absorptivity spectra for bare InAs and the GMR structure at $\theta_i = 50°$.
Fig. 5: Intensity effect of GMR on nonreciprocal absorption at $\theta_i = 70°$.
Fig. 6: Magnetic field strength and angular dependence of reciprocity breaking.

## Supplementary Materials

Fig. S1: Slab waveguide angular dispersion for varying $\alpha$-Si layer thicknesses on top of $n$-InAs.
Fig. S2: Schematic of reflection setup and confirmation of the Onsager reciprocal relations.
Fig. S3: Simulated and experimental data on the narrow-angle transition from strong-to-weak nonreciprocal absorption.
Fig. S4: Experimental and simulated spectra for the absorptivity at varying incident angles and magnetic field strengths.
Fig. S5: Temperature dependence of the dielectric constant of InAs.
Fig. S6: Electric field intensity plots showing the plasmon mode confinement within the GMR structure.
Figure S7: Image and schematic of the Halbach array with tunable supermendur pole pieces.

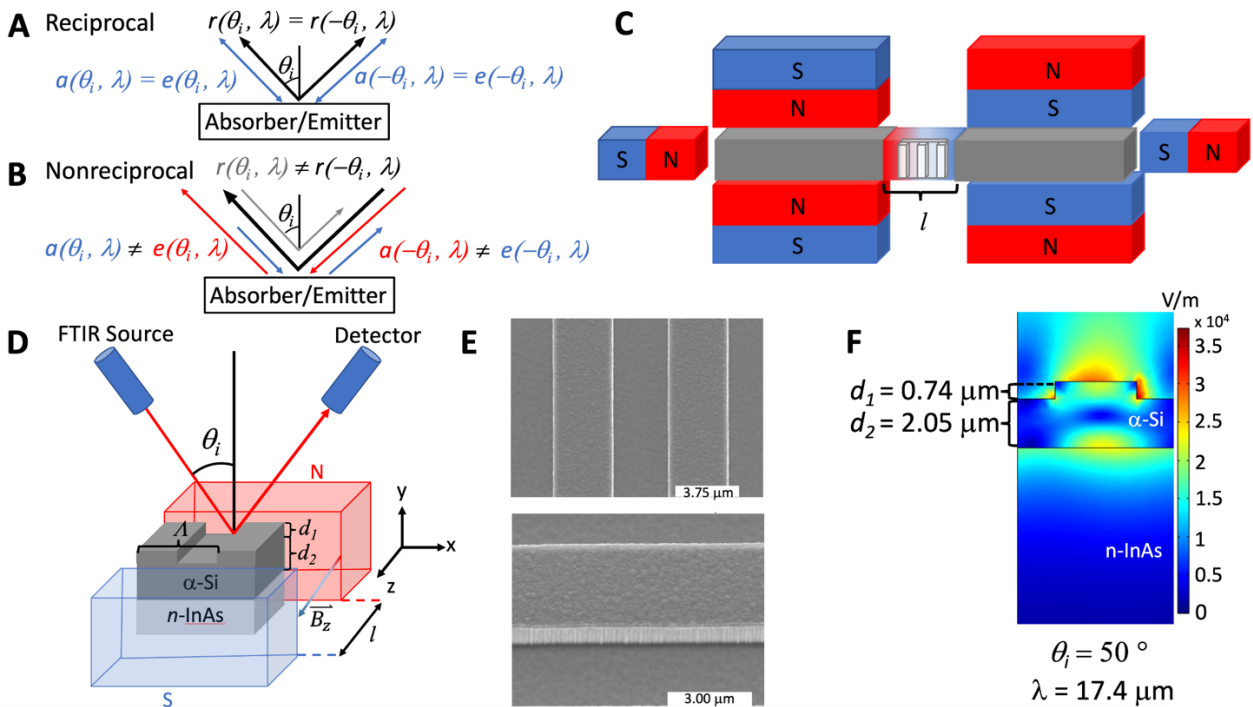

**Fig. 1: Overview of nonreciprocal absorber/emitter theory, design, and implementation. (A & B)** Reciprocal and nonreciprocal relations for an absorber/emitter. **(C)** Schematic of the Halbach array used in the measurement. The poles of the magnets on each side of the system are rotated by 90° relative to one another to increase the magnitude of the magnetic field through the pole pieces (gray). The pole pieces provide a focused and uniform magnetic field across the sample, which is shown in the gap. We tune the magnetic field strength by changing the gap length, $l$, between the pole pieces. **(D)** Schematic of the measurement scheme. A silicon carbide Globar is used as the thermal source inside of a Fourier transform infrared (FTIR) spectrometer. The sample is mounted on a rotating stage which controls the angle of incidence, $\theta_i$, from the source onto the sample. The detector is mounted on a rotating arm to collect the specular reflected light. Polarizers at the source output and detector input allow polarization-dependent measurements. **(E)** Scanning electron microscope (SEM) images of the α-Si photonic crystal slab with no tilt (top), and 45° tilt (bottom). **(F)** Simulation showing the electric field intensity for 50° incident radiation at 17.3 μm with the measured α-Si parameters from the SEM images.

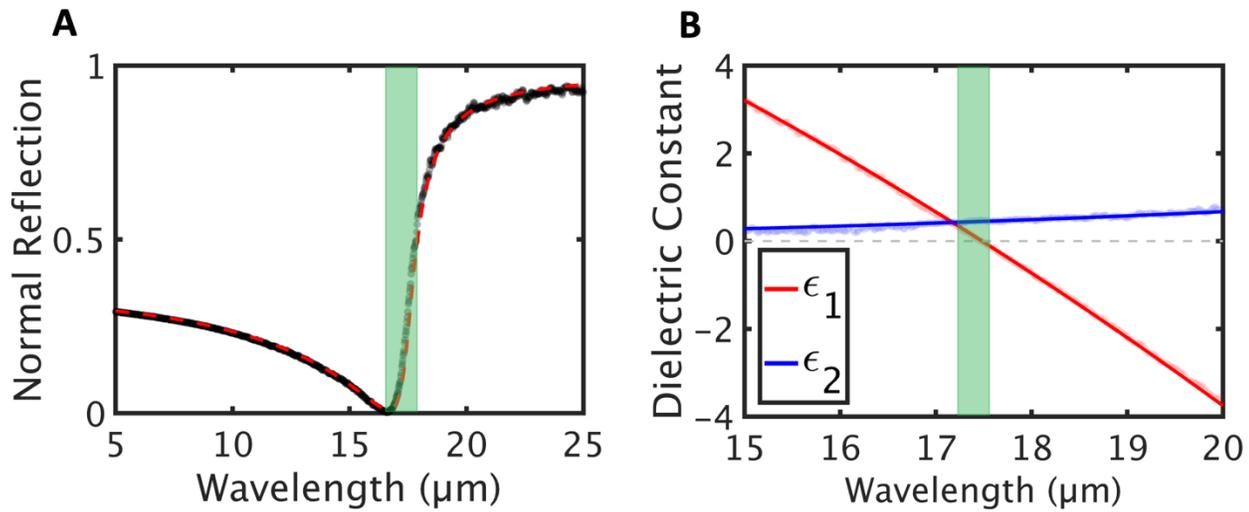

**Fig. 2: Drude reflectance and dielectric constant of degenerately-doped *n*-InAs.** (**A**) Drude reflectance of the InAs wafer. The transparent black dots show the measured data and the dashed red curve gives the fit for a Drude-like optical response with $n = 1.5 \times 10^{18} \, \text{cm}^{-3}$, and $\Gamma = 4.5$ THz. (**B**) Real and imaginary parts of the dielectric function for the isotropic case. The transparent dots mark the permittivity values directly calculated from the data, and the solid blue and red curves show the extracted values from fitting the ellipsometry values $\Psi$ and $\Delta$. The chartreuse strips mark ENZ wavelength regime where we expect to see large nonreciprocal behavior.

$\theta_i = 50°$

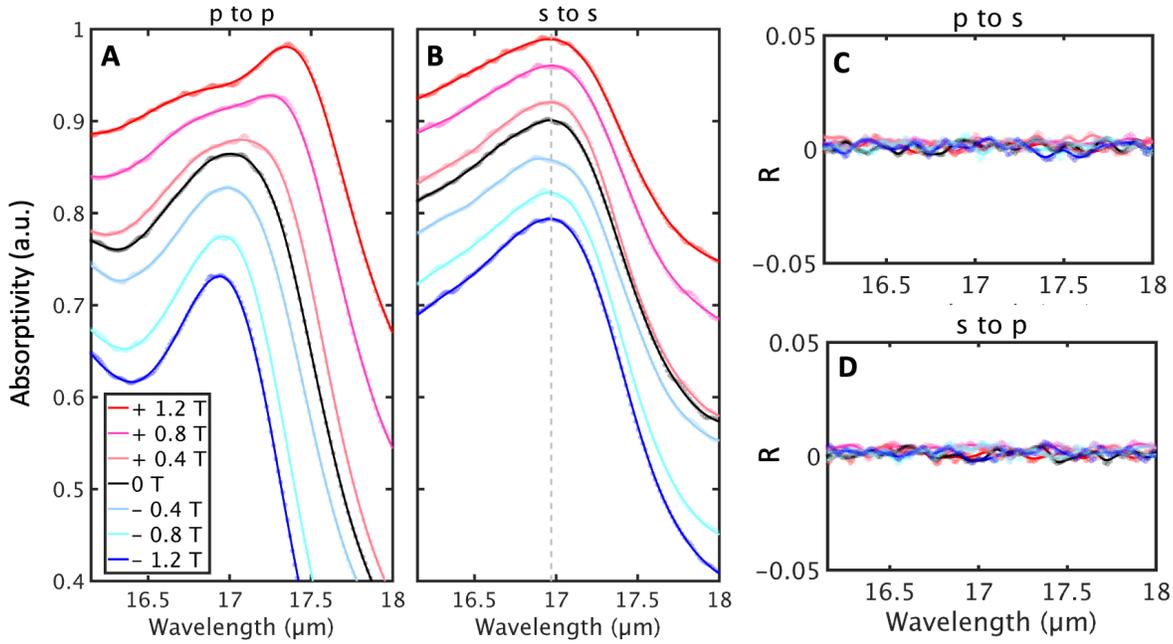

**Fig. 3: Polarization-dependent reflection measurements for varying applied magnetic field**.
(**A**) *p* to *p*-polarization, (**B**) *s* to *s*-polarization, (**C**) *s* to *p*-polarization, and (**D**) *p* to *s*-polarization.
The dotted points are experimental data and the solid lines are fits to the data points. We only
show data for the incident angle $\theta_i = 50°$ in this figure; data for other incident angles included in
the SI. We see a clear tuning of the peak position with magnetic field for *p* to *p*-polarization but
no spectral shape change for *s* to *s*-polarization. We do not observe polarization conversion,
further demonstrating that the change in the *p* to *p*-polarization is not due to misalignment of the
device in the setup. The data in (A) and (B) are artificially offset to highlight the spectral peak
shift (*p*-polarization) and no shift (*s*-polarization). The data for other angles for *p*-polarization
(Figures S2-4) also shows an intensity change that overall aligns with simulations.

$$\theta_i = 50°$$

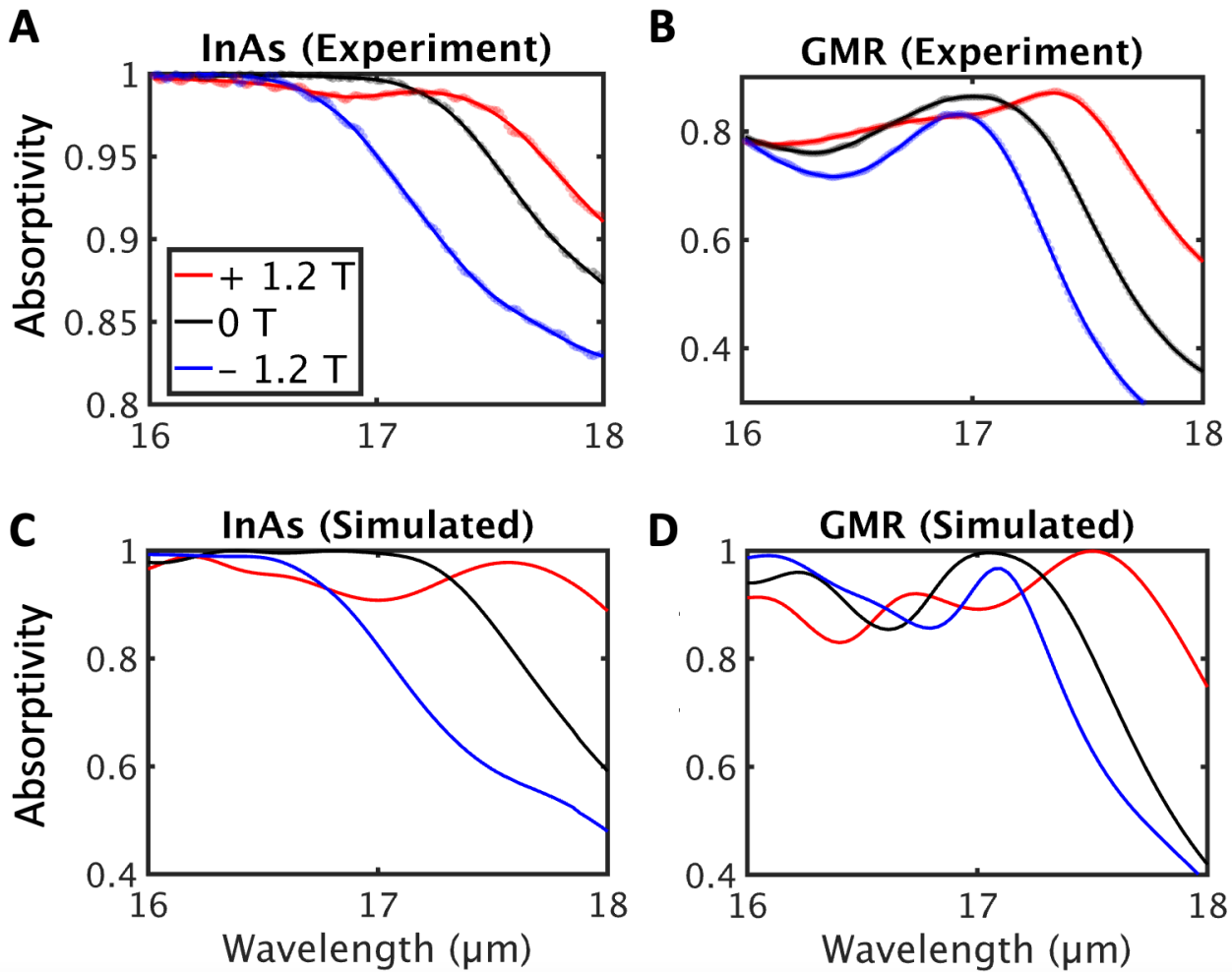

**Fig. 4: Magnetic field effect on *p*-polarized absorptivity spectra for bare InAs and the GMR structure at $\theta_i$ = 50°.** (**A**) The measured shift in absorptivity of bare InAs as a function of magnetic field emanates from the splitting of the plasma edge in InAs into a positive (red) and negative (blue) magneto-plasmon. (**B**) Data for the GMR structure, showing the effect of adding the α-Si structure on top of the InAs wafer, which produces resonant absorptivity peaks for both positive and negative applied magnetic fields. Panels **C** and **D** show the simulated results for bare InAs and the GMR structure, respectively.

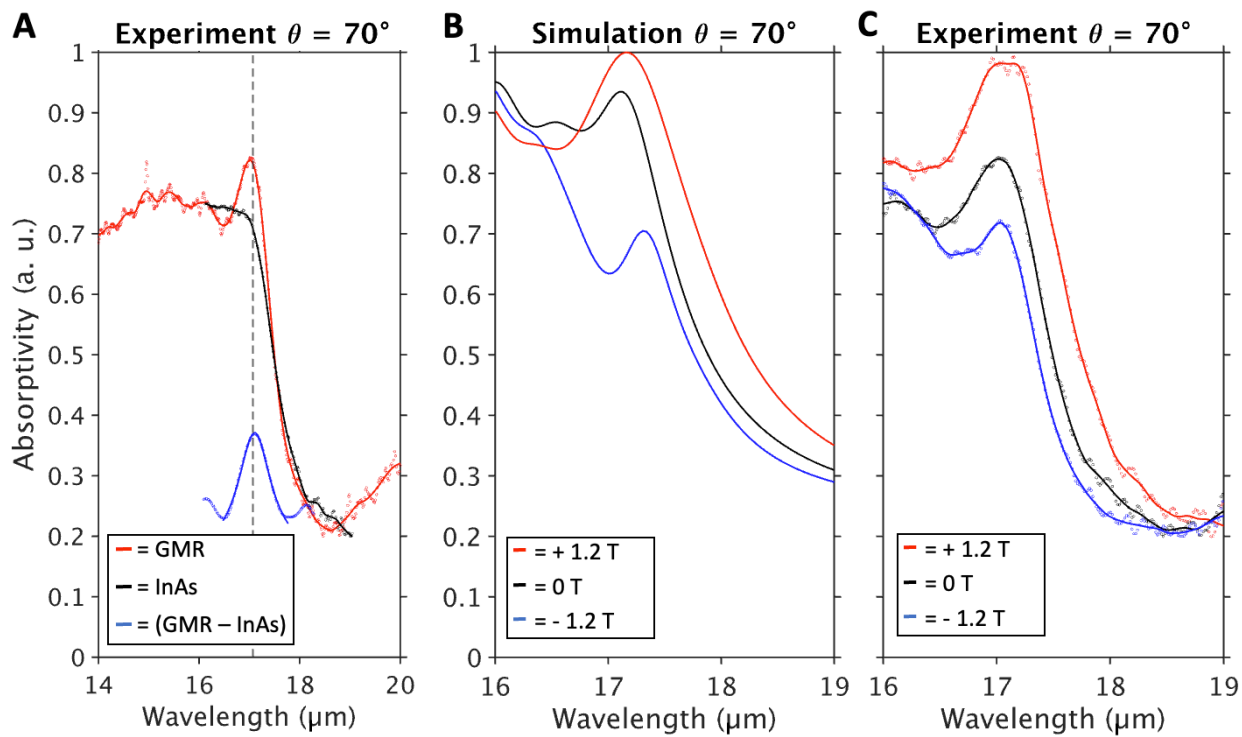

**Fig. 5: Intensity effect of GMR on nonreciprocal absorption at $\theta_i = 70°$.** (**A**) Experimental data at 70° incidence for the GMR structure and *n*-InAs with no patterned α-Si on top. The blue trace is the offset difference between the GMR and the *n*-InAs to highlight the resonant effect of the GMR. (**B** and **C**) Simulated and experimental data for positive and negative field showing the strong resonance intensity tuning at this angle.

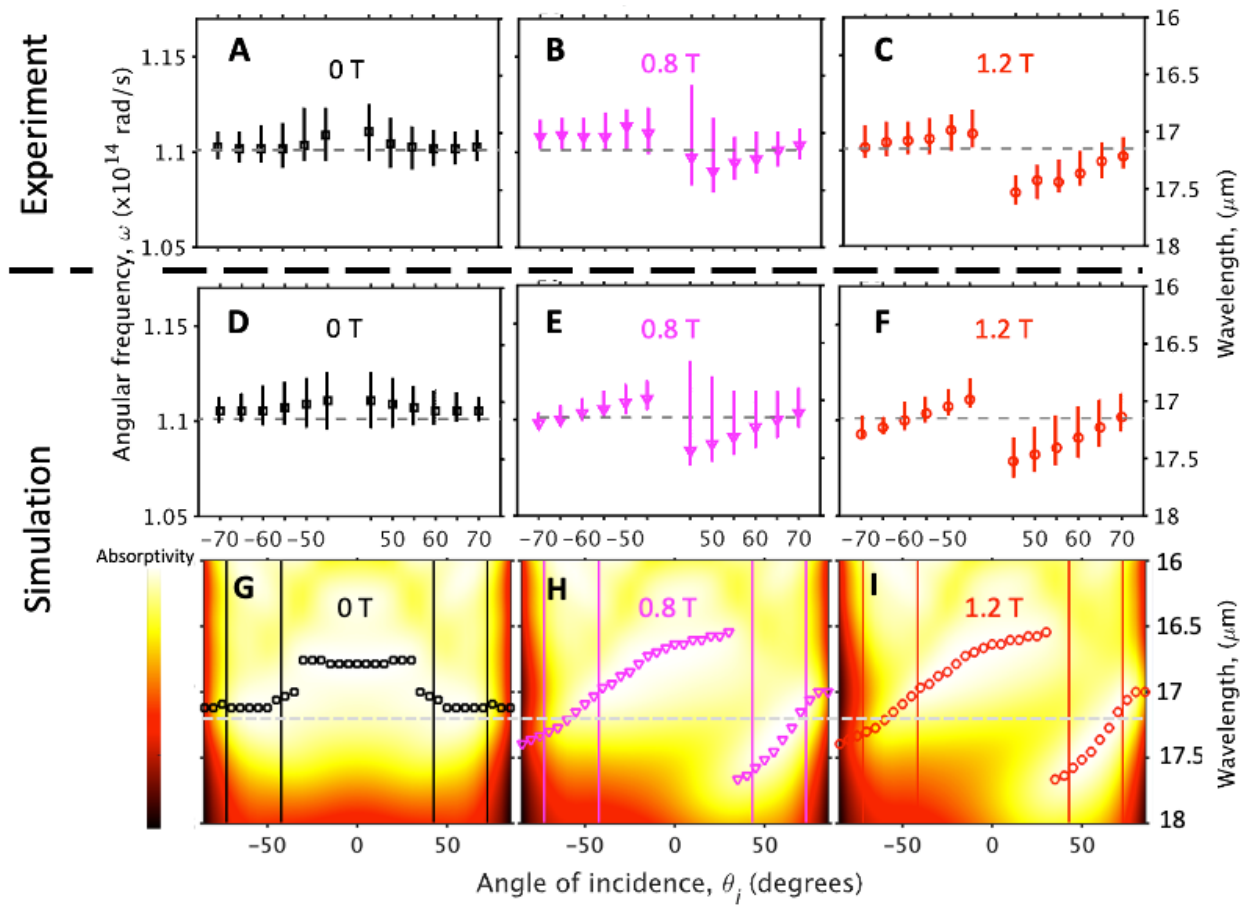

**Fig. 6: Magnetic field strength and angular dependence of reciprocity breaking.** (**A-C**) Compiled experimental and simulated (**D-F**) absorptivity maxima for *p*-polarized light as a function of magnetic field and angle of incidence. When no magnetic field is applied (**A & D**) the structure behaves reciprocally. At 0.8 T (**B & E**) and 1.2 T (**C & F**) the *n*-InAs no longer satisfies Lorentz reciprocity and the reciprocal absorptivity relationship is broken. The error bars represent data points in the spectra to within 5% of the maximum value to give the reader an idea of the line shape. Plots of the individual spectra for other angles are included in the Supplementary Material (Figures S2-4). (**G-I**) Heatmap of the simulated absorptivity over the entire angle of incidence range for varying magnetic fields. Horizontal lines mark the positive (and negative) incidence ranges shown in panels **A-F**.